\documentclass[runningheads]{llncs}
\usepackage{physics}
\usepackage{amsmath,amssymb}
\usepackage{subcaption}
\usepackage{amssymb}
\usepackage{graphicx}
\usepackage{wrapfig}
\usepackage[linesnumbered,ruled,vlined]{algorithm2e}
\usepackage{upquote}
\title{Q-means using variational quantum feature embedding}
\author{Arvind S. Menon\thanks{arvind6599@gmail.com} \& Nikaash Puri\thanks{nikpuri@adobe.com}}
\institute{Media and Data Science Research Lab, Adobe India}

\begin{document}
%

\maketitle
%

\date{May 2021}
\begin{abstract}
This paper proposes a hybrid quantum-classical algorithm that learns a suitable quantum feature map that separates unlabelled data that is originally non linearly separable in the classical space using a Variational quantum feature map and q-means as a subroutine for unsupervised learning. The objective of the Variational circuit is to maximally separate the clusters in the quantum feature Hilbert space. First part of the circuit embeds the classical data into quantum states. Second part performs unsupervised learning on the quantum states in the quantum feature Hilbert space using the q-means quantum circuit. The output of the quantum circuit are characteristic cluster quantum states that represent a superposition of all quantum states belonging to a particular cluster. The final part of the quantum circuit performs measurements on the characteristic cluster quantum states to output the inter-cluster overlap based on fidelity. The output of the complete quantum circuit is used to compute the value of the cost function that is based on the Hilbert-Schmidt distance between the density matrices of the characteristic cluster quantum states. The gradient of the expectation value is used to optimize the parameters of the variational circuit to learn a better quantum feature map. 
\end{abstract}
\section{Introduction}
\begin{figure}[!htb]
    \centering
    \begin{minipage}{0.49\textwidth}
        \centering
        \includegraphics[width=1\textwidth]{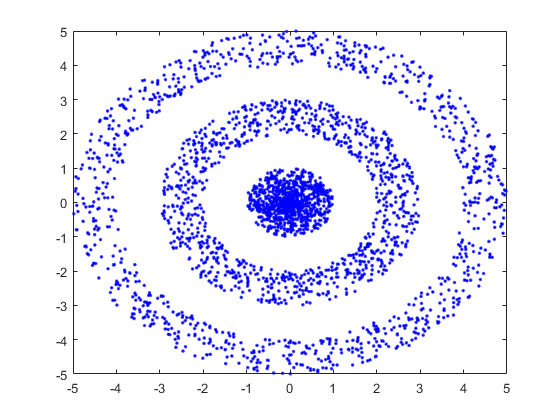} \\
         (a)Cartesian co-ordiantes \\(all clusters have the same mean)
    \end{minipage}\hfill
    \begin{minipage}{0.49\textwidth}
        \centering
        \includegraphics[width=1\textwidth]{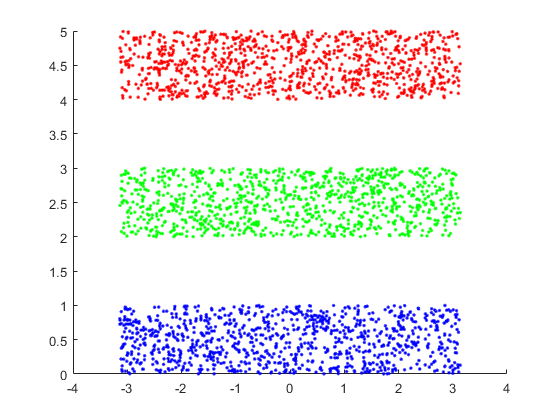}\\
         (b)Polar co-ordinates
    \end{minipage}
    \caption{Linear separability of the clusters in polar space}
    \label{fig:figl}
\end{figure}
The quantum algorithm Q-means \cite{qmeans} theorises an exponential speedup over the conventional K-means classical machine learning algorithm. However, it performs cluster assignment based on Euclidean distances, meaning that it would not work for non-linearly separable clusters (Fig.~\ref{fig:figl}) in the original space, e.g. Concentric circles, Moons, Corners etc. Thus there is a need to implement feature mapping / kernel methods to deal with data in a higher dimensional space making data highly separable, giving better clustering results.

The idea of “quantum metric learning” discussed in \cite{lloyd2020quantumembedding} Quantum embedding provides an efficient way to train a variational circuit to find a hyperplane that separates the labelled data in the quantum Hilbert feature space. It uses the labels provided alongside the data to find an efficient quantum feature map to perform supervised learning. \cite{schuld2018QFHS} The idea of encoding classical data into quantum states can be interpreted as a quantum feature map, which can help in finding the underlying the patterns within the unlabelled data. The variational circuit can be used to perform quantum computations that are classically intractable to obtain classically intractable feature maps in a high dimensional quantum feature Hilbert space.

After the data is embedded into the quantum Hilbert space, Q-means can be implemented on the quantum states to obtain characteristic cluster quantum states i.e. the quantum states that are superposition all the quantum states in a particular cluster. A cost function based on the cluster quantum states can be evaluated by measuring the overlap or fidelity between the characteristic cluster quantum states, ensuring high separation between clusters and low cluster scatter about the centroids. The optimization of the cost function is done using the gradient of the quantum circuit operations. This gradient is evaluated using the automatic differentiator discussed in \cite{bergholm2018pennylane}. Subsequently the quantum parameters for the variational feature map are updated for a better quantum feature map.

Dealing with non-trivial clustering tasks requires kernels that deal in higher dimensions where it is highly likely to be linearly separable.

NISQ Devices have been successfully used to implement Kernel methods in supervised learning \cite{lloyd2013supunsup,bartkiewicz2019experimentalkernal,blank2019tailoredkernel,havlicek2018enhanced,schuld2018QFHS,Schuld_2017,afham2020quantumknn,li2019sublinear}  to obtain promising results using different type of classifiers such as kernel-based, $\ell^{2}$ - margin SVM, minimum enclosing ball, K-Nearest Neighbours, distance based classifiers etc .\cite{havlicek2018enhanced} suggests two methods for supervised classification, first based on variational circuits to learn the function to approximate a classifier and second being a quantum SVM with a quantum feature map to implement a classically intractable kernel function for unique results.\cite{QCL} discusses the application of approximating any teacher function given a labelled dataset by training a a Variational circuit to build a classifier. A quantum circuit can use a mixture of both ideas for unsupervised learning that learns a classically intractable quantum circuit for the input data.

The classical problem of Kernel K-means is an integer programming problem making it an NP-Hard problem. Other approaches such as adiabatic computing have been used as another method for K-means to find the best labelling possible for the encoded data, discussed in \cite{bauckhage2017adiabatic}. The problem is encoded into a Hamiltonian and the quantum state evolves with time towards the ground state of the problem Hamiltonian. The result is a quantum state that has the highest amplitude for the best possible labelling solution.

Finding an optimal solution to the k-means clustering problem for d-dimensional observations is an NP-Hard problem in euclidean space, even in the case of binary clustering. If we fix k and d values then the complexity of finding an optimal solution is $\mathbb{O}(n^{kd+1})$

The current work deals with supervised learning, which reduces the complexity and changes the nature of the problem. The purpose of the paper is to learn a quantum metric to find a pattern in the unlabelled data that consists of non-convex clusters.

This will eliminate the problem of finding a suitable kernel that works best for the dataset. As we know in the classical case there is no one fits all kernel that outputs optimal clustering solutions for all kinds of datasets. Leveraging the universal nature of QAOA in quantum computation, we try to use the high dimensionality of the Hilbert space to find an optimal quantum feature map that are classically intractable.

Section 2 illustrates the methodology used to implement the algorithm. Section 3 illustrates the complete algorithm using pseudocode. Section 4 discusses the observations based on the simulations and inferences based on the results.  Section 5 is a discussion regarding further possible applications of this classical-quantum hybrid algorithm. The final section of the paper is the conclusion.

\section{Methodology}
\subsection{Variational Quantum feature map $U(x,\theta)$}
\begin{figure}[t]
    \centering
    \includegraphics[width=\linewidth]{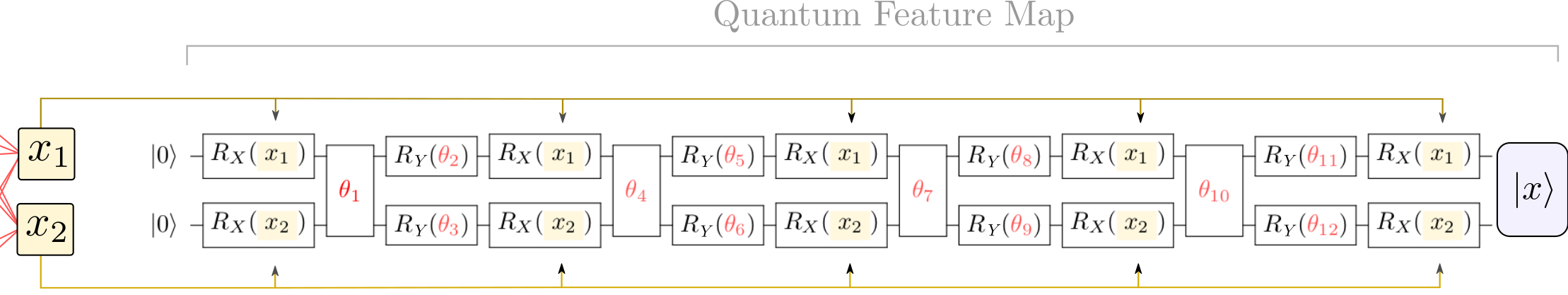}
    \caption{Quantum variational circuit implementing the quantum feature map \cite{lloyd2020quantumembedding} giving $\ket{x}$ referred to as $U(x,\theta)$ for a two dimensional data point (x1,x2).}
    \label{fig:fig2}
\end{figure}

From the perspective of quantum computing, a
quantum feature map $x \mapsto \ket{x}$ corresponds to
a quantum circuit $U(x,\theta)$ that takes classical data as input and performs feature encoding by applying quantum gates or unitary transformations on a ground or vacuum state $\ket{0...0}$ of a Hilbert space\cite{schuld2018QFHS}  $\mathcal{F}$ to produce quantum states:
\begin{equation}
    \ket{x} = U(x,\theta)\ket{00...0}
\end{equation}
The choice of feature map is very important to the algorithm. This decides the nature of the kernel function $\mathbb{K}(x_{i},c_{j})$= $|\bra{x_{i}}\ket{c_{j}}|^{2}$.
First part of the quantum circuit involves encoding unlabelled classical data using quantum gates (Fig.~\ref{fig:fig2}) that perform unitary operations on the initial quantum state based on quantum parameters $\theta$ and input data. A variational circuit is used to perform quantum feature encoding to learn the best possible quantum feature map $U(x,\theta)$  to obtain meaningful clustering results.
\paragraph{Advantage of using a Variational quantum feature map :\\}
$U(x,\theta)$ explores the best representation space to find non-trivial patterns underlying the data in a quantum feature Hilbert space.
Also, the quantum feature Hilbert space dimensions are exponential with respect to qubits used in the initial state. This way high dimensional feature spaces can be efficiently implemented.
The ansatz proposed in \cite{lloyd2020quantumembedding} also used in this paper is based on the QAOA framework. Repetitions of this ansatz can implement classically intractable feature maps based on the universality of QAOA in quantum computing\cite{lloyd2018quantumQAOA}.

\begin{wrapfigure}{l}{0.3\textwidth}
\includegraphics[width=\linewidth]{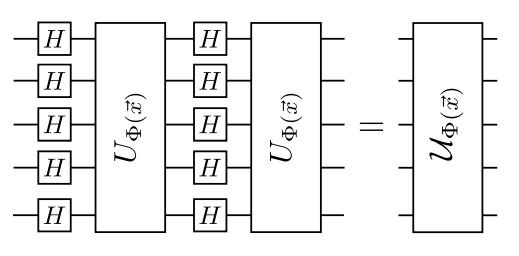}
\caption{\cite{havlicek2018enhanced}quantum circuit for the general circuit $U_{\phi(x)}$}
\label{fig:fig3}
\end{wrapfigure}
\subsubsection{Fixed state preparation}
 If we chose a feature map such that $\mathbb{K}(\Vec{x_{i}},\Vec{c_{j}})$= $|\bra{x_{i}}\ket{c_{j}}|^{2}$ is too simple then the quantum circuit is unable to provide a quantum advantage. An example of a  simple output from a feature map quantum circuit would be product states as it is easily simulated classically. Since quantum circuits are hard to simulate classically there does exist a list of quantum circuits that could do the required operations, providing a quantum advantage. 

An example of such a quantum circuit is discussed in \cite{havlicek2018enhanced}, which suggests a feature map on n-qubits as $U(x,\phi) = U_{\phi(x)}H^{\otimes n}U_{\phi(x)}H^{\otimes n}$, where H is the standard Hadamard gate and $U_{\phi(x)}$ is :
\begin{equation}
U_{\phi(x)}= exp\left( i \sum_{s\subset [n]}\phi_{S}(x) \prod_{i\in S} Z_{i} \right)    
\end{equation}

\begin{figure}[t]
    \centering
    \includegraphics[width=0.7\linewidth]{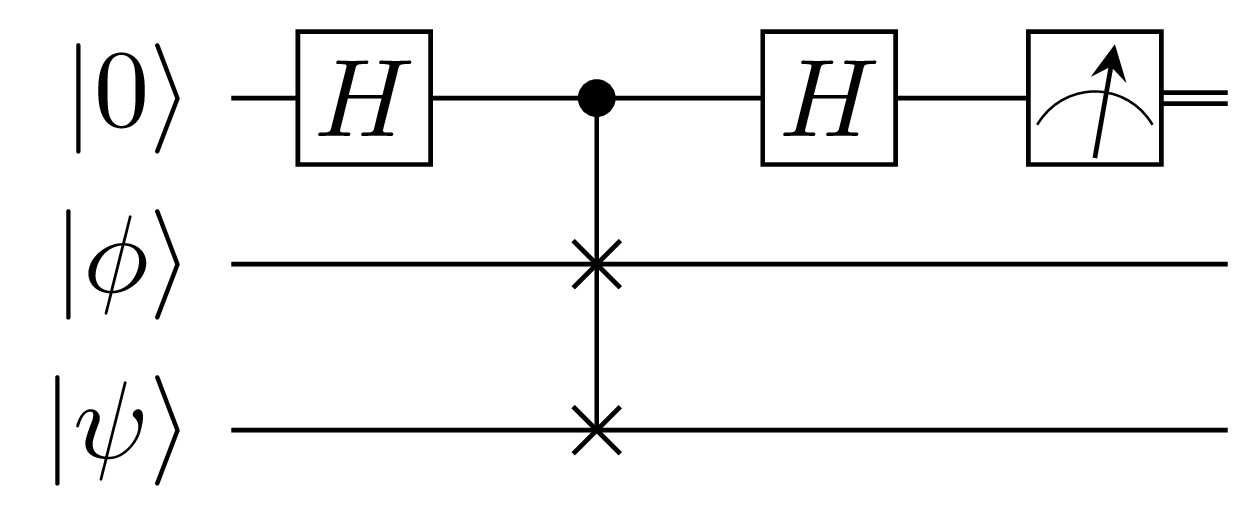}
    \caption{A swap test using ancilla qubits. The final measurement of the ancilla qubit in the Pauli-Z basis performed multiple times yields the estimate of probability of measuring $\ket{0}$ which is proportional to the kernel function value.}
    \label{fig:fig4}
\end{figure}

\subsection{Q-means: Kernel estimation and labelling based on swap test}
An unsupervised algorithm that performs K-means on quantum states to output centroid quantum states based on the unlabelled input data assumed to be stored in a Q-RAM \cite{qmeans}. The proposed algorithm makes use of q-means in a slightly different way. Q-means will be used a subroutine to output characteristic cluster quantum states $\ket{\chi_j}, j \in [1-k]$. 
\begin{equation}
    \ket{\chi_j} = \frac{1}{\sqrt{|C_j|}}\sum_{x_{i} \in C_{J}}\ket{x_{i}}
\end{equation}
This form of output is necessary because of the nature of the measurement to be made on the output of the q-means based subroutine.  

\subsubsection{Kernel estimation} Perform the mapping:
\begin{equation}
\frac{1}{\sqrt{N}}\sum_{i=1}^{i=N}\ket{i}\otimes_{j\in [K]}\ket{j}\ket{0} \mapsto \frac{1}{\sqrt{N}}\sum_{i=1}^{i=N}\ket{i}\otimes_{j\in [K]}\ket{j} \ket{\overline{\mathbb{K}(x_{i},c_{j})}}
\end{equation}

where $\mathbb{K}(x_{i},c_{j} = |\bra{x_i}\ket{c_{j}}|^{2}$ and $|\overline{\mathbb{K}(x_{i},c_{j})} - \mathbb{K}(x_{i},c_{j}) | \leq \epsilon_{1}$\\
Details of this step are illustrated in Kernel estimation section.

\subsubsection{Cluster membership assignment} The max value out of all the K register containing the kernel value is used to assign the cluster membership and generate labels for each data point\cite{durr1996quantumin}.

\begin{equation}
 \frac{1}{\sqrt{N}}\sum_{i=1}^{i=N}\ket{i}\otimes_{j\in [K]}\ket{j} \ket{\overline{\mathbb{K}(x_{i},c_{j})}} \mapsto \frac{1}{\sqrt{N}}\sum_{i=1}^{i=N}\ket{i}\ket{l_{i}} 
\end{equation}

\subsubsection{Characteristic cluster quantum states}
On measuring the label register, the first qubit collapses to a superposition of data points belonging to the observes cluster label. This way we are able to obtain the characteristic cluster quantum state $\ket{\chi_j} = \frac{1}{\sqrt{|C_j|}}\sum_{x_{i} \in C_{J}}\ket{i}\ket{x_{i}}$

\subsubsection{Cluster centroid update}
Matrix multiplication of matrix $X^{T}$ and vector $\ket{\chi_j}$ to obtain the state $\ket{c_{j}^{t+1}}$ with error $\epsilon_{2}$, along with an estimation of $ |\ket{c^{t+1}_{j}}| $ with relative error $\epsilon_{3}$

Once the convergence condition is satisfied for q-means subroutine, the $\ket{\chi_{j}}$ are measured in the next section of the quantum circuit to evaluate the cost function.

\subsubsection{Method of kernel estimation and Labelling}
As illustrated in Fig~\ref{fig:fig2}, computing of the kernel function starts by applying $U(x, \theta)$ and hadamard gate to the ancillary qubit :
\begin{equation}
\frac{1}{\sqrt{N}}\sum_{i=1}^{i=N}\ket{i}\otimes_{j\in [K]}\ket{j}\ket{0}_{a}\ket{0}\ket{0} \mapsto \frac{1}{\sqrt{N}}\sum_{i=1}^{i=N}\ket{i}\otimes_{j\in [K]}\ket{j}\frac{\left(\ket{0}+\ket{1} \right)}{\sqrt{2}} \ket{x_{i}}\ket{c_{j}}
\end{equation}
Now, swapping is done between the data point and the centroid qubits using ancilla as the control quibit to obtain the state. 
\begin{equation}
\frac{1}{\sqrt{N}}\sum_{i=1}^{i=N}\ket{i}\otimes_{j\in [K]}\ket{j}\frac{1}{\sqrt{2}}\left(\ket{0}\ket{x_{i}}\ket{c_{j}}+\ket{1}\ket{c_{j}}\ket{x_{i}}\right)
\end{equation}
Once again, Hadamard gate is applied to the accilary qubit to create interfernce between the swapped qubits to obtain the final state ($ \psi^{\pm}_{ij}=\ket{x_{i}}\ket{c_{j}} \pm \ket{c_{j}}\ket{x_{i}}$) :
\begin{equation}
\frac{1}{\sqrt{N}}\sum_{i=1}^{i=N}\ket{i}\otimes_{j\in [K]}\ket{j}\frac{1}{2}\left(\ket{0}\ket{\psi^{+}_{ij}}+\ket{1}\ket{\psi^{-}_{ij}}\right) 
\end{equation}
The probability of obtaining $\ket{0}$ when the third register is measured is,
\begin{equation}
p_{0}=\bra{\psi^{+}}\ket{\psi^{+}}=\frac{1+|\bra{x_{i}}\ket{c_{j}}|^{2}}{2}= \frac{1 + \mathbb{K}(x_{i},c_{j})}{2}
\end{equation}
\cite{qmeans} Rewriting the final state by swapping the last two registers ($\ket{0}\ket{\psi^{+}} \mapsto \ket{\psi_{ij},0}$ (G is a garbage state)  we can summarize the whole algorithm as :
\begin{equation}
 \mathcal{A}: \ket{i}\ket{j}\ket{0}\ket{0}\ket{0}\mapsto \ket{i}\ket{j}\left( \sqrt{p_{0}}\ket{\psi_{ij},0}+\sqrt{1-p_{0}}\ket{G,0}\right)
\end{equation}

Now that we know how to apply the transformation $\mathcal{A}$ defined in Eqn. 10,  \cite{brassard2000quantum} amplitude estimation can be used to create the desired state, which stores the kernel value based on Theorem :
\begin{theorem}
Given algorithm $\mathcal{A}$,for any positive integer P, the amplitude estimation algorithm
outputs $\hat{p_{0}} (0 \leq \hat{p_{0}} \leq 1)$ such that :
\begin{equation}
    |\hat{p_{0}}-p_{0}| \leq 2\pi\frac{\sqrt{p_{0}(1-p_{0})}}{P}+(\frac{\pi}{P})^{2}
\end{equation}
with probablity 8/$\pi^{2}$. It uses exactly P iterations of the algorithm A. If $p_{0} = 0$ then $\hat{p_{0}} = 0$
with certainty, and if $p_{0} = 1$ and P is even, then $\hat{p_{0}} = 1$ with certainty.
\end{theorem}

There are recent developments in amplitude estimation \cite{AEwithoutPE} that which requires lesser circuit depth focusing on implementation on NISQ devices by avoiding phase estimation.
\\
A simple final step of finding the maximum $p_{0}$ value to label ($j^{*}$) each data point, as the highest kernel value implies the greatest similarity to a particluar cluster's centroid quantum state. \cite{durr1996quantumin} can be used to find the lowest $\hat{p_{1}}$ which is inversely proportional to the Kernel value as the amplitudes are normalized. 
\begin{equation}
    \j^{*} = arg max_{j \in \{1,..k\}}\{\hat{p_{0}}\}
\end{equation}
where $j^{*}$ is the label for the data point $\Vec{x_{i}}$

\paragraph{Advantage of using q-means as a subroutine for unsupervised learning}:\\ Data is readily available in the form of quantum states after the quantum feature map performs quantum embedding. Implying that the input does not not have to be interpreted as classical data using conventional feature encoding such as basis encoding or amplitude encoding, and has more freedom as it already lies in the high dimensional quantum feature Hilbert space. It is not restricted by a certain method of state preparation thus finding the local optimum kernel that performs well for the encoded data.

The algorithm effectively uses quantum parallelism for distance calculation and labelling each data point, which is the primary reason for the exponential speed-up in it's complexity over the classical counterpart.

\paragraph{Limitations of the q-means subroutine} Neither quantum simulators nor
quantum computers large enough to test q-means are available currently as discussed in \cite{qmeans}. Although the results of Q-means are consistent with the results of $\delta$-K means which is a noisier version of the conventional K-means. With advancement in quantum technology in the future q-means will be implementable on a quantum computer.

\begin{figure}[t]
\begin{subfigure}{.5\textwidth}
  \centering
  \includegraphics[width=\linewidth]{Quantum-swap-test-circuit-correct.png}
  \label{fig:sfig1}
\end{subfigure}%
\begin{subfigure}{0.5\textwidth}
  \centering
  \includegraphics[width=1\linewidth]{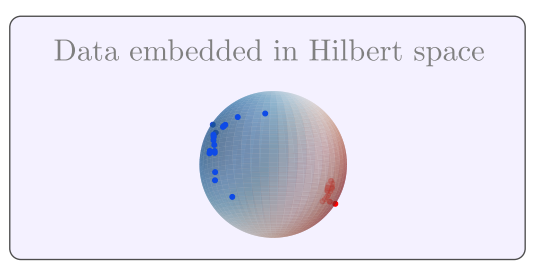}
  \label{fig:sfig2}
\end{subfigure}
\caption{left) A swap test to measure the overlap between the characteristic cluster quantum states $\ket{\phi}$ and $\ket{\psi}$ right)Representation of 2 clusters $\ket{\psi}$ and $\ket{\phi}$ on a bloch sphere \cite{lloyd2020quantumembedding} }
\label{fig:fig5}
\end{figure}
\subsection{Measurement}
The previous section outputs the set $\{\ket{\chi_{j}}: j \in [1-k]\}$. This section which forms the final part of the quantum circuit (Fig.~\ref{fig:fig4}) consists of a measurement that computes the overlap between the characteristic cluster quantum states. This is used as a measure of the separation between the clusters. The swap test calculates the fidelity between the quantum states.\\
\begin{equation}
    \mathbb{M}_{jj'} = |\bra{\chi_{j}}\ket{\chi_{j'}}|^{2}
\end{equation}



\subsection{Cost function and $\Vec{\theta}$ update}
All the measurements representing the expectation value of the observable are summed up to form a cost $C(\theta)$ function representing the total inter-cluster overlap in the quantum feature Hilbert space.
\begin{equation}
    C(\theta) = \sum_{j\ne j'}\mathbb{M}_{jj'} = \sum_{j\ne j'}|\bra{\chi_{j}}\ket{\chi_{j'}}|^{2}  \text{  $\forall$ j $\in$ [1-k]}
\end{equation}
This cost function is similar to the Hilbert Schmidt distance ($D_{hs})$ used in \cite{lloyd2020quantumembedding}. $D_{hs}$ is used to quantify the underlying distance between the clusters, which depends on the overlap between the quantum states embedded in each cluster. If k=2 and $\ket{a}$ is from cluster A and $\ket{b}$ belongs to cluster B then $D_{hs}$ is given by  :
\begin{equation}
    D_{hs}(A,B) = \frac{1}{2}\left(\sum_{i,i'}|\bra{a_{i}}\ket{a_{i'}}|^{2} + (\sum_{j,j'}|\bra{b_{j}}\ket{a_{j'}}|^{2}\right) - \sum_{i,j}|\bra{a_{i}}\ket{b_{j}}|^{2}
\end{equation}
and the respective cost function is simply C = 1 - $D_{hs}(A,B)$

Since q-means uses data in superposition to obtain an exponential speed-up and lesser qubit usage, it also leads to restrictions in the nature of the cost function. Restrictions such as inability to calculate intra-cluster scatter as $|\bra{\chi_{0}}\ket{\chi_{0}}|^{2} = 1$. The resultant cost function thus reformulates the objective as minimizing the overlap between the clusters.

The quantum parameter $\theta$ is updated using the gradient of the $C(\theta)$ which is computed using Pennylane automatic differentiator \cite{bergholm2018pennylane}. The RMSPropOptimizer along with the QAOAEmbedding framework was used to run simulation illustrated in Fig~\ref{fig:fig5}.

\section{Algorithm}
The psuedocode illustrates the framework for the Variational q-means based on all the steps that were described in detail in the previous sections.

\begin{algorithm}
\KwIn{Random initialization of $\theta$ for the Variational quantum circuit. Choose hyperparameter \textbf{step size} for the Pennylane RSMPropOptimizer and error parameters $\epsilon_{1},\epsilon_{2},\epsilon_{3},\epsilon_{4},$.The data matrix $X\in\mathbb{R}^{n\times d}$ stored in a QRAM. Initialize K centroids $c^{0}_{0},c^{0}_{1},...c^{0}_{K}$ using K-means ++ method, and store it in the QRAM. K ancilla qubits each initialized to $ \ket{0}$ to compute fidelity. }
\KwOut{Overlap cost function and trained parameter $\overline{\pmb{\theta}}$ representing the learnt quantum feature map $U(x,\overline{\pmb{\theta}})$}

initial iteration t=0\linebreak
\textbf{Quantum computation}\\
\textbf{Part 1: Variational Quantum Feature Map U(x,$\theta$)}\linebreak
\textbf{Feature embedding} 
\[
\ket{x} = U(x,\theta)\ket{00..0}
\]
\textbf{Create a superposition state} Using an oracle to form a superposition state based on the index of the data point.
\[
\frac{1}{\sqrt{N}}\sum_{i=1}^{i=N}\ket{i}\ket{x_{i}}\otimes_{j\in [K]}\ket{j}\ket{c_{j}}\ket{0}_{a}
\]
\\
\textbf{Part 2: Q-means subroutine}\linebreak
\textbf{Kernel Estimation}
\[
\frac{1}{\sqrt{N}}\sum_{i=1}^{i=N}\ket{i}\otimes_{j\in [K]}\ket{j}\ket{\mathbb{K}(\Vec{x_{i}},\Vec{c_{j}})}
\]
where, $\mathbb{K}(x_{i},c_{j})$= $|\bra{x_{i}}\ket{c_{j}}|^{2}$.\linebreak
\textbf{Cluster Assignment}
\[
\frac{1}{\sqrt{N}}\sum_{i=1}^{i=N}\ket{i}\otimes_{j\in [K]}\ket{j}\ket{\mathbb{K}(\Vec{x_{i}},\Vec{c_{j}})} \mapsto \frac{1}{\sqrt{N}}\sum_{i=1}^{i=N}\ket{i}\ket{l^{t}(v_i)}
\]
\textbf{Characteristic cluster quantum state and Centroid estimation}
Label quantum register is measured to obtain state $\ket{\chi'_j} = \frac{1}{\sqrt{|C_j|}}\sum_{x_{i} \in C_{J}}\ket{i}$ with probability $\frac{|C_j|}{N}$. Matrix multiplication of matrix $X^{T}$ and vector $\ket{\chi_j}$ to obtain the state $\ket{c_{j}^{t+1}}$
$t=t+1$,\\
Repeat \textit{Steps 3 to 6}, till $|c_{j+1}-c{j} |<\epsilon_{3}$.\\
Compute $\{\ket{\chi_j} : \forall j \in [1-k] \}$ using the final $\ket{c_{j}^{t+1}}$ quantum states.\\
\textbf{Part 3: Measurement}\linebreak
\textbf{Compute cluster overlap based on fidelity}
\[
\mathbb{M}_{jj'} = |\bra{\chi_{j}}\ket{\chi_{j'}}|^{2}
\]
 All inter-cluster overlaps are measured using swap test between each pair of distinct characteristic cluster quantum state.\linebreak
\textbf{Classical computation}\\
Compute cost function $C(\theta)$  based on the output of the quantum circuit. 
\[
C(\theta) = \sum_{j\ne j'}\mathbb{M}_{jj'} = \sum_{j\ne j'}|\bra{\chi_{j}}\ket{\chi_{j'}}|^{2}
\]
Update quantum parameter $\theta$ to obtain a lower cost function.\\
Repeat the process till $|C^{t+1}(\theta)-C^{t}(\theta) |<\epsilon_{4}$.

\caption{Variational Quantum Kernel K-means}
\label{algo 1}
\end{algorithm}

\begin{figure}[!htb]
    \centering
    \begin{minipage}{0.4\textwidth}
        \centering
        \includegraphics[width=1.1\textwidth]{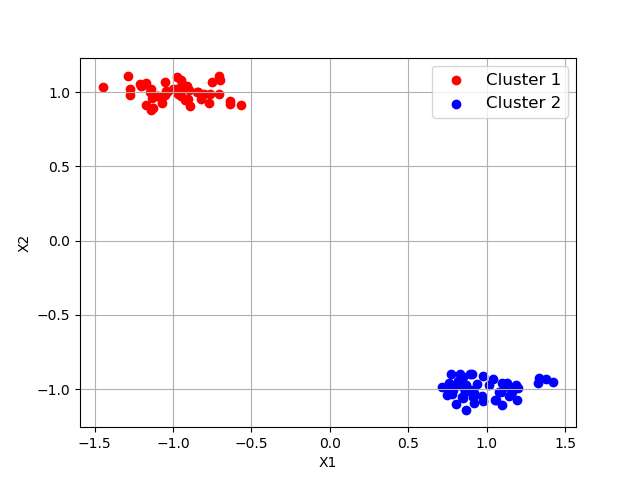} \\
    \end{minipage}\hfill
    \begin{minipage}{0.6\textwidth}
        \centering
        \includegraphics[width=\textwidth]{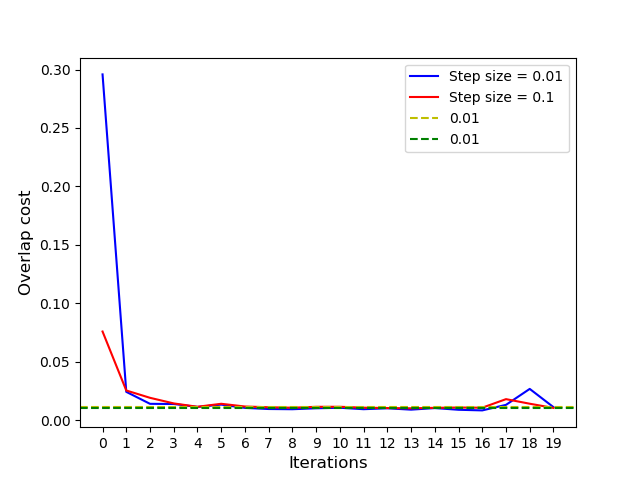}\\
    \end{minipage}
    \begin{minipage}{0.4\textwidth}
        \centering
        \includegraphics[width=1.1\textwidth]{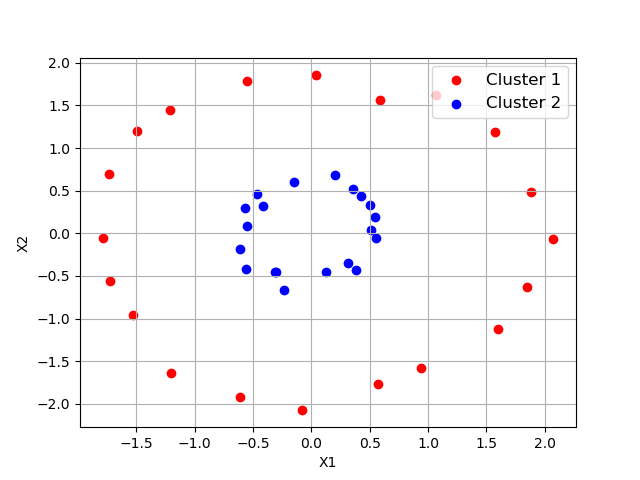} \\
    \end{minipage}\hfill
    \begin{minipage}{0.6\textwidth}
        \centering
        \includegraphics[width=\textwidth]{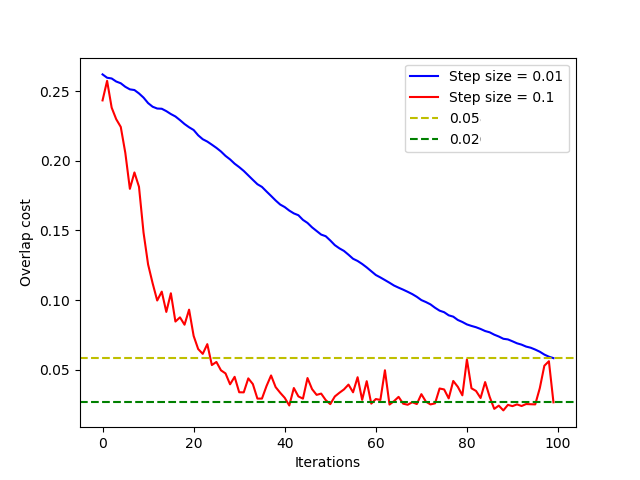}\\
    \end{minipage}
    \label{fig:fig6}
        \begin{minipage}{0.4\textwidth}
        \centering
        \includegraphics[width=1.1\textwidth]{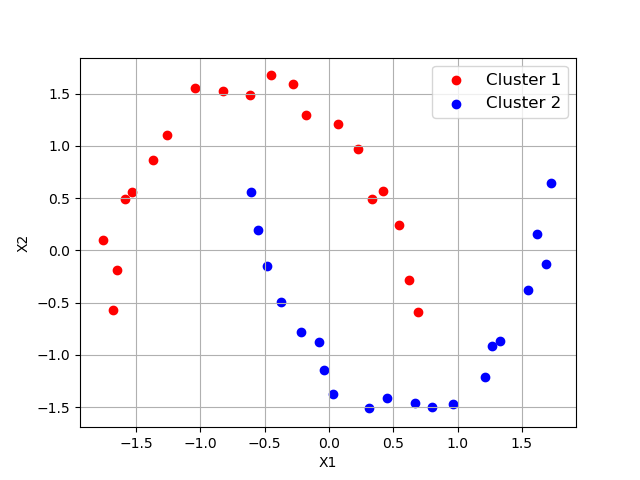} \\
        (a)Clusters in original space
    \end{minipage}\hfill
    \begin{minipage}{0.6\textwidth}
        \centering
        \includegraphics[width=\textwidth]{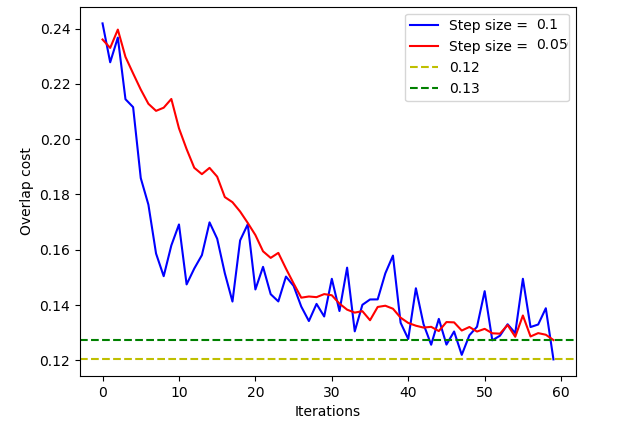}\\
         (b) $C(\theta)$ vs Iterations for different step sizes
    \end{minipage}
    \caption{Convergence of overlap cost function for various data sets}
    \label{fig:fig6}
\end{figure}
\section{Results from simulations}
The simulations (Fig~\ref{fig:fig6}) to compute overlap cost function are implemented using Pennylane's QAOA embedding framework$\footnote{Pennylane: Quantum embedding and metric learning}$.All the datasets are pre-processed using Sklearn before quantum embedding. Currently no quantum simulators are available to test q-means, so cluster datasets with labels were used as inputs for the quantum embedding circuit to train the quantum partameters.

\begin{table}[h!]
\centering
\begin{tabular}{|c|c|c|c|} 
\hline
Dataset & Min. $C(\theta)$ value & Iteration of min value & Better Step size \\ [0.5ex] 
\hline
Blobs & 0.01  & $4^{th}$ & 0.1 \\
Concentric circles & 0.02 & $34^{th}$ & 0.1 \\
Moons & 0.08& $154^{th}$ & 0.15\\
\hline
\end{tabular}
\caption{Simulation results}
\end{table}
\paragraph{Observations} In Fig~\ref{fig:fig6}, the blobs dataset is very well separated in the original space, due to which the algorithm did not take many iterations to converge. Whereas in the case of non-linearly separable datasets, the algorithm took more number of iterations to converge. The quantum parameters had to be trained to find the pattern that is non-trivial in the classical space. Step size = 0.1 is observed to perform better than the other step sizes. 

\section{Discussion}
\paragraph{Evaluating clustering performance using quantum metrics}

Clustering performance of an unsupervised learning can be evaluated using the overlap measurement to quantify the separation between the clusters in the quantum Hilbert space. The output labeled data can be mapped using quantum embedding and learn a good feature mapping, then score it based on the measurement - based distance between the density matrices using $\ell$-1 or $\ell$-2 norms.

\paragraph{Interpretation of the quantum parameters} The trained parameter $\theta$ represents the  representation space learnt by the algorithm for the embedded dataset. The final quantum feature map can be used to interpret underlying pattern in the dataset.

\section{Conclusion}
Despite a complete understanding of the idea of clustering discussed in \cite{impossible}, this paper aims to create maximal separation between dissimilar quantum states to perform clustering. The algorithm learns a quantum feature map by training quantum parameters based on labels obtained from the q-means unsupervised learning subroutine.  The repetitive approach of performing unsupervised learning followed by adaptively training the feature map aims at learning the best representation of the data in the quantum feature Hilbert space. By explicitly implementing the kernel estimation, we are able to explore more feature spaces that are better suited for the embedded dataset. Since the current existing simulators cannot be used to implement the complete algorithm, the quantum advantage is yet to be experimentally shown.

%
%
\bibliographystyle{splncs04}
\bibliography{references}

\end{document}